\documentstyle{europhys}

\def\etal{{\hbox{{\xpt\it\ et al.\/}\xpt\rm :\ }}}

\def\And{{\rm and\ }}

\def\drm{{\rm d}}

\def\stars{\bigskip\centerline{***}\medskip}

\newif\ifboo \boofalse

\def\Review#1{\boofalse{\it #1},}
\def\Name#1{{\sc #1},}
\def\Vol#1{\ifboo Vol. {\bf #1}\else{\bf #1}\fi}
\def\Year#1{\ifboo #1\else(#1)\fi}
\def\Book#1{\bootrue{\it #1},}
\def\Page#1{\ifboo {\rm p. #1}\else{\rm #1}\fi}

\begin{document}
\euro{}{}{}{} 
\Date{Version of July 2, 1999}
\shorttitle{K. BRODERIX \etal SHEAR VISCOSITY OF A CROSSLINKED POLYMER MELT}
\title{Shear viscosity of a crosslinked polymer melt} 
\author{Kurt Broderix\inst{1,2}, 
        Henning L\"owe\inst{1}, 
        Peter M\"uller\inst{1} \And 
        Annette Zippelius\inst{1}}
\institute{
  \inst{1} Institut f\"ur Theoretische Physik - Universit\"at
           G\"ottingen, D-37073 G\"ottingen, Germany\\
  \inst{2} Department of Physics, Theoretical Physics - University of Oxford,
           Oxford OX1 3NP, United Kingdom}
\rec{}{} 
\pacs{ \Pacs{64}{60.Ht}{Dynamic critical phenomena}
\Pacs{82}{70.Gg}{Gels and sols} 
\Pacs{83}{10.Nn}{Polymer dynamics} }
\maketitle
\begin{abstract}
  We investigate the static shear viscosity on the sol side of the
  vulcanization transition within a minimal mesoscopic model for the
  Rouse-dynamics of a randomly crosslinked melt of phantom polymers.
  We derive an exact relation between the viscosity and the
  resistances measured in a corresponding random resistor network.
  This enables us to calculate the viscosity exactly for an ensemble
  of crosslinks without correlations. The viscosity diverges
  logarithmically as the critical point is approached. For a more
  realistic ensemble of crosslinks amenable to the scaling description
  of percolation, we prove the scaling relation $k=\phi-\beta$ between
  the critical exponent $k$ of the viscosity, the thermal exponent $\beta$
  associated with the gel fraction and the crossover exponent $\phi$
  of a random resistor network.
\end{abstract}
\section{Introduction}
As the gelation or vulcanization transition is approached from the sol side, a
melt or solution of polymers becomes increasingly more viscous, suggesting a
divergence of the static shear viscosity $\eta$ at the critical point.  The
transition is commonly interpreted as a signature of a percolation transition
\cite{StCoAd82,DaLa90,Sa94}: When the concentration $c$ of crosslinks, which
bind different polymers together, is increased to its critical value $c_{\rm
  crit}$ a macroscopic cluster of polymers is formed. Even for chemical
gelation, {\it i.e.}\ permanent crosslinking, the measured values of the
exponent $k$ of the viscosity $\eta\sim(c_{\rm crit}-c)^{-k}$ lie in a broad
range.  Adam {\it et al.}\ \cite{AdDeDuHiMu81-AdDeDu85} find values $0.6\leq
k\leq 0.9$ for polycondensation without solvent and radical copolymerisation
with solvent.  Finite frequency measurements by Durand {\it et al.}\ 
\cite{DuDeAdLu87} also yielded results within the above range. Silica gels and
epoxy resins have been investigated by Martin {\it et al.}\ 
\cite{MaWi88-AdMa90-MaWiOd91,MaAdWi88} and by Colby {\it et al.}\ 
\cite{CoCoSaMe87,CoGiRu93}.  The results are in the range $1.1\leq k\leq 1.7$.
For a review see \cite{MaAd91}.

Using the percolation picture a number of heuristic proposals have been made as
to how the exponent $k$ is related to the critical exponents arising in the
scaling description \cite{StAh94} of percolation.  The most common proposal,
$k=2\nu-\beta$, has firstly been given by de~Gennes \cite{Ge78} within a Rouse
approximation. Here $\nu$ is the exponent governing the divergence of the
correlation length and $\beta$ is associated to the gel fraction. This result
has been re-derived and supported with additional arguments in
\cite{MaAdWi88,ArSa90,CoGiRu93,Sa94}. On the other hand, the underlying
assumptions have been questioned \cite{StCoAd82}.  In \cite{Ge79} de~Gennes
hinted at an analogy between the viscosity and the conductivity of a random
mixture of superconductors and normal conductors, see \cite{StCoAd82} for
details. The resulting conjecture is $k=s$, where $s$ is the exponent ruling
the divergence of the conductivity. In \cite{Ke83} Kert\'esz argued for
$s=\nu-\beta/2$ in analogy to the Alexander-Orbach conjecture \cite{AlOr82}.
This is, at least, a reasonable approximation for $s$. The resulting value for
$k$ happens to be exactly the one suggested in \cite{ArSa90,Sa94} for Zimm
dynamics as opposed to Rouse dynamics. In fact, \cite{ArSa90,Sa94} point out
that the factor two by which their proposals for $k$ for the two dynamics in
question differ fits nicely to the two experimentally observed regimes.

As can be seen most clearly from the arguments given in
\cite{CoGiRu93}, the basis for a representation of $k$ in terms of the
exponents of percolation theory is to relate the longest relaxation
time of a cluster of macromolecules to the cluster's size. This
relaxation time, of course, does not only depend on the assumed
dynamical model but on the internal structure of the cluster as well.
The internal structure of
percolation clusters is by now believed to be characterized by
an exponent independent of $\beta$ and $\nu$, the latter
ruling the behaviour of the clusters on large scales. This
believe stems from the failure of the Alexander-Orbach
conjecture \cite{AlOr82}, see \cite{HaLu87} and
references therein.  More quantitatively, the spectral
dimension $d_s$ of the incipient spanning cluster introduced
in \cite{AlOr82} is only approximately but not exactly equal
to $4/3$ for spatial dimensions $d<6$. The spectral dimension
encodes the fractal nature of the cluster's connectivity
without giving reference to its spatial configuration. This
can be seen, for example, from the fact that $d_s$
parametrizes the Lifshits tail $D(E)\sim E^{(d_{s}/2)-1}$ of
the density of states $D(E)$ of the discrete Laplacian within
the cluster.
Cates \cite{Ca85} relates the
relaxation spectrum of a connected fractal cluster of $n$ polymers to
$D(E)$ and finds that the static shear viscosity within Rouse
approximation scales as $n^{(2/d_{s})-1}$.

Since the different heuristic arguments yield competing proposals even
for virtually the same set of assumptions, the appropriate way is to
discuss clear cut models.  The purpose of this Letter is threefold:
\quad (i)~~Following the classical Kirkwood approach \cite{DoEd85} we
first calculate the static shear viscosity of a monodisperse sol of
phantom monomer chains, which follow Rouse dynamics and are
permanently crosslinked by Hookean springs chosen at random
\cite{SoVi95}.  Identifying the crosslinks with electrical resistors
soldered together by the polymers, the polymer network may be thought
of as a random resistor network.  This allows us to establish an
\emph{exact} correspondence between the viscosity and the resistance
of a random resistor network.  \quad (ii)~~For the simplest
distribution of crosslinks without any correlations, which amounts to
a mean-field like model of percolation, this correspondence is
employed to derive an exact expression for the averaged viscosity,
implying a logarithmic divergence at the gelation transition.
  \quad (iii)~~For general distributions of
crosslinks, assumed to be amenable to the scaling description of
percolation, the exact correspondence between the viscosity and the
resistance is employed to establish the scaling relation
\begin{equation} \label{Eq1}
  k=\phi-\beta\,.
\end{equation}
The exponent $\phi$ was first introduced in the context of random
resitor networks. It governs the growth of the resistance ${\cal
  R}(r)$ between two points on the incipient spanning cluster, which
are a large spatial distance $r$ apart: ${\cal R}(r)\sim r^{\phi/\nu}$
\cite{HaLu87,LuWa85,StJaOe99}.  The exponent $\phi$ is related to the
spectral dimension, according to  $\phi= \nu
d_{f}\bigl((2/d_{s})-1\bigr)$, where $d_{f}=
d-\beta/\nu$ is the Hausdorff-Besicovitch dimension of the fractal.

\section{Dynamic Model} 
We consider a system of $N$ linear, identical, mono-disperse polymer chains,
each consisting of $L$ monomers.  Monomer $s$ on chain $i$ is characterized by
its time-dependent position vector ${\bf R}_t(i,s)$ ($i=1,\ldots,N$ and
$s=1,\ldots,L$) in $d$-dimensional space. We are interested in shear flows and
impose an external velocity field $v^{\alpha}_{{\rm ext}}({\bf r},t)$. Here,
Greek indices indicate Cartesian co-ordinates, {\it i.e.}\
$\alpha=x,y,z,\dots$. We will always use a flow field in the $x$-direction,
increasing linearly with $y$, {\it i.e.}\ $v^{x}_{{\rm ext}}({\bf r},t) =
\kappa(t)y$ and $v^{\alpha}_{{\rm ext}}({\bf r},t)=0$ for $\alpha\neq x$. We
employ the simplest, purely relaxational dynamics \cite{DoEd85}
\begin{equation} \label{Eq2}
  \partial_t R^{\alpha}_t(i,s) = 
  - \frac{1}{\zeta} \: \frac{\partial H}{\partial R^{\alpha}_t(i,s)} 
  + v^{\alpha}_{{\rm ext}}\bigl({\bf R}_t(i,s),t\bigr) + \xi^{\alpha}_t(i,s).
\end{equation}
In the course of time, the monomers relax to the state $\partial H/\partial{\bf
  R}={\bf 0}$ such that their velocity is equal to the externally imposed
velocity field, {\it i.e.}\ $\partial_t R^{\alpha}_t(i,s)=v^{\alpha}_{{\rm
    ext}} \bigl({\bf R}_t(i,s),t\bigr)$. The relaxation process is disturbed by
thermal noise {\boldmath$\xi$} with zero mean and covariance
$\overline{\xi^{\alpha}_t(i,s)\,\xi^{\beta}_{t'}(i',s')} = {2}{\zeta}^{-1}\,
\delta_{\alpha,\beta}\, \delta_{i,i'}\, \delta_{s,s'}\, \delta(t-t')$.  Here,
the overbar indicates the average over the realizations of the Gaussian noise
{\boldmath$\xi$}.  The relaxation constant is denoted by $\zeta$, and we use
energy units such that $k_B T=1$. In (\ref{Eq2}) the Hamiltonian $H:=H_W+U$ is
supposed to consists of two terms. The first one
\begin{equation} \label{Eq3}
  H_W := \frac{d}{2l^2}\sum_{i=1}^{N}\sum_{s=1}^{L-1} 
  \bigl({\bf R}(i,s+1)-{\bf R}(i,s)\bigr)^2
\end{equation}
guarantees the connectivity of each chain, the typical distance between
monomers being given by the persistence length $l>0$.  The second one models
$M$ permanently formed crosslinks between randomly chosen pairs of monomers
${\cal G}:=\{i_e,s_e;i_e',s_e'\}_{e=1}^M$. To constrain the relative distance
between two monomers, participating in a crosslink, we choose a harmonic
potential
\begin{equation} \label{Eq4}
  U := \frac{d}{2a^2}\:\sum_{e=1}^M 
  \bigl( {\bf R}(i_e,s_e)-{\bf R}(i'_e,s'_e) \bigr)^2,
\end{equation}
whose strength is controlled by the parameter $a>0$. For $a\to 0$ hard
crosslinks can be recovered \cite{SoVi95}.  Since the Hamiltonian is quadratic
in the monomer positions, it can be expressed in terms of a $NL\times NL$
random connectivity matrix $\Gamma$ according to
$H=:(d/2a^{2})\sum_{i,i'=1}^{N}\sum_{s,s'=1}^{L} {\bf R}(i,s)\cdot
\Gamma(i,s;i',s')\,{\bf R}(i',s')$.
\section{Shear viscosity and resistor networks}
Our aim is the computation of the intrinsic shear stress
$\sigma_{\alpha\beta}(t)$ as a function of the shear rate $\kappa(t)$.
Following \cite{DoEd85}, we express the shear stress in terms of the force per
unit area, exerted by the polymers
\begin{equation} \label{Eq5}
  \sigma_{\alpha\beta}(t) =
  -\frac{\rho_{0}}{N}\sum_{i=1}^{N}\sum_{s=1}^L 
  \overline{F^{\alpha}_t(i,s) R^{\beta}_t(i,s)}.
\end{equation}
Here, $\rho_{0}$ stands for the polymer concentration and
$F^{\alpha}_t(i,s):=-\partial H/\partial R^{\alpha}_t(i,s)$ is the force on
monomer $(i,s)$.  Since the dynamic equation (\ref{Eq2}) is linear in the
monomer positions ${\bf R}_{t}$, it can be readily solved.  Upon inserting this
solution into (\ref{Eq5}), we find the linear relation
\begin{equation}\label{Eq6} 
  \sigma_{xy}(t) = 
  \int_{-\infty}^{t} \drm\tau \: G(t-\tau) \kappa(\tau), \qquad
  G(t) := 
  \frac{\rho_{0}}{N}\; {\rm Tr}
  \biggl((1-E_0)\exp\biggl\{-\,\frac{2dt\Gamma}{\zeta a^{2}}\biggr\}\biggr),
\end{equation}
where $E_{0}$ is the projector onto the null space
of the connectivity matrix $\Gamma$.  For a constant shear rate $\kappa$, the
intrinsic viscosity $\eta$ is related to the stress tensor via
$\eta:=\sigma_{xy}/(\kappa\rho_{0})$ such that
\begin{equation} \label{Eq7}
  \eta({\cal G}) = 
  \rho_{0}^{-1}\int_0^{\infty}\!\drm t \: G(t) = 
  \frac{\zeta a^{2}}{2dN}\:{\rm Tr}\!\left(\frac{1-E_0}{\Gamma}\right) 
\end{equation}
is given by the trace of the Moore-Penrose inverse \cite{Al72} of $\Gamma$,
{\it i.e.}\ the inverse of $\Gamma$ restricted to the subspace of non-zero
eigenvalues. Note that we have made the dependence of $\eta$ on the realization
${\cal G}$ of the crosslinks explicit.

Each crosslink realisation ${\cal G}$ defines a random labelled graph, which
can be decomposed into its maximal path-wise connected components or clusters,
${\cal G}=\cup_{k=1}^{K}{\cal N}_{k}$. The associated connectivity matrix is of
block-diagonal form so that
\begin{equation} \label{Eq8}
  \eta({\cal G}) = \sum_{k=1}^{K} \frac{N_{k}}{N}\:\eta({\cal N}_{k}).
\end{equation} 
Here, $N_k$ denotes the number of polymers in the cluster ${\cal N}_{k}$.

Let us identify a bond between two neighbouring monomers on the same polymer as
a resistor of magnitude $l^{2}/a^{2}$ and a crosslink between polymers as a
resistor of magnitude $1$.  The resistance measured between any
connected pair of vertices $(i,s)$ and $(i',s')$ will be denoted by
${\cal R}(i,s;i',s')$. 
From (\ref{Eq7}) and \cite[Thm.~F]{KlRa93} it follows that
\begin{equation} \label{Eq9}
  \eta({\cal N}_k) =  
  \frac{\zeta a^{2}}{4d L N_k^2} \sum_{(i,s;i',s')\in{\cal N}_k}
  {\cal R}(i,s;i',s'). 
\end{equation} 
Together with (\ref{Eq8}) this constitutes the announced 
connection between the
viscosity and the resistances in a random resistor network.

We expect an observable, like the viscosity, to be self-averaging in the
macroscopic limit and compute the average $\langle\eta\rangle$ over all
crosslink realisations in the macroscopic limit $N\to\infty, M\to\infty$ with
the concentration of crosslinks $c:=M/N$ being fixed. It will be advantageous
to reorder the sum in (\ref{Eq8}) by summing first over all clusters consisting
of given number $n$ of polymers and subsequently over all ``sizes'' $n$.  Thus
we obtain for the averaged viscosity
\begin{equation} \label{Eq10}
  \left\langle\eta\right\rangle 
  = \sum_{n=2}^{\infty} n\tau_{n} \left\langle\eta\right\rangle_{n},
\end{equation}
where $\langle\eta\rangle_{n} := \tau_{n}^{-1} \bigl\langle
N^{-1}\sum_{k=1}^{K} \delta_{N_{k},n}\,\eta({\cal N}_{k})\bigr\rangle$ denotes
the average of $\eta$ over all clusters of size $n$ and $\tau_{n} := \langle
N^{-1}\sum_{k=1}^{K} \delta_{N_{k},n}\rangle$ denotes the average number of
clusters of size $n$ per polymer.
\section{Completely random crosslinks}
The simplest distribution of crosslinks omits correlations between crosslinks
and gives equal weight to all possible crosslink realisations.  This case has
been studied extensively in the theory of random graphs, as developed by
\cite{ErRe60}. More precisely, we will calculate
$\langle\eta\rangle := \lim_{N\to\infty} \bigl( \prod_{e=1}^{cN}
(NL)^{-2} \sum_{i_e,i'_e=1}^{N} \sum_{s_e,s'_e=1}^{L}\bigr) \eta(\{i_e, s_{e};
i_e', s'_{e}\})$ in this section. In the macroscopic limit there are no
clusters of macroscopic size for $c<c_{{\rm crit}}:=\frac{1}{2}$, and all
monomers belong to tree clusters without loops \cite[Thms.~5d,e]{ErRe60}.
Moreover \cite[eq.\ (2.18)]{ErRe60}, one has
$\tau_{n}=n^{n-2}(2c\,{\rm e}^{-2c})^n/(2c\,n!)$. Since all $L^{2(n-1)}n^{n-2}$
trees ${\cal T}_{n}$ of size $n$ can be proven to occur equally likely, 
the average with respect to them reduces to $\langle\eta\rangle_{n} = 
\linebreak[1] L^{2(1-n)}n^{2-n}\sum_{{\cal T}_{n}} \eta({\cal T}_{n})$.
Note further that the resistance between two vertices in a tree simplifies
considerably, because there is a unique path connecting them, implying that all
resistors are in series. In the simplest case of crosslinked Brownian
particles, {\it i.e.}\ polymers consisting of just one monomer each ($L=1$),
the resistance ${\cal R}(i;i')$ is equal to the number of crosslinks connecting
$i$ and $i'$. Hence, we refer to \cite[Thm.~1]{MeMo70} for computing 
\begin{equation} \label{Eq11}
  \langle {\cal R}(i;i')\rangle_{n} = 
  (n-2)!\,\sum_{\nu=1}^{n} \frac{n^{1-\nu}\,\nu(\nu -1)}{(n-\nu)!}
  \stackrel{n\to\infty}{\simeq} \sqrt{n\pi/2}, \quad i\neq i'.
\end{equation}
The double series arising from (\ref{Eq10}), (\ref{Eq9}) and (\ref{Eq11}) can
be summed up in closed form so that we obtain the averaged viscosity of
crosslinked Brownian particles for all $0<c<\frac{1}{2}$. This result is
readily generalised to networks of crosslinked polymer chains ($L\ge 1$). The
only difference is that now there are two different kinds of resistors with
magnitudes $l^{2}/a^{2}$ and $1$, corresponding to intra-polymer bonds and
crosslinks, respectively. Due to the fact that the monomer labels $s$ are
distributed independently from the chain labels $i$, one can reduce the
averaged resistance to the Brownian-particle case (\ref{Eq11}). This gives rise
to the exact result
\begin{equation} \label{Eq12}
  \left\langle \eta\right\rangle  =  
  \frac{\zeta a^2}{8cd} \left( L + \frac{l^{2}}{3a^{2}}(L^{2}-1) \right)
  \left[ \ln\!\left(\frac{1}{1-2c}\right) - 2c \right] 
   + \frac{\zeta l^{2}}{12d}(L^{2}-1)
\end{equation}
for the averaged viscosity of a crosslinked polymer melt on the sol side
$0<c<\frac{1}{2}$. It exhibits a logarithmic divergence at the critical
concentration $c_{{\rm crit}}=\frac{1}{2}$ corresponding to the critical
exponent $k=0$. The result (\ref{Eq12}) is universal in the sense that the
details of the model only affect the pre-factor of the critical divergence,
which depends on the persistence length $l$, the ``extension'' of the
crosslinks $a$ and the length $L$ of a polymer chain. For long polymers the
viscosity is proportional to $L^2$, as it should be within a Rouse-type model
\cite{DoEd85}. This scaling is not altered when passing to the limit
$a\downarrow 0$ of hard crosslinks. 

The disorder average of the viscosity has also been computed
\cite{Broderix99} with the
help of the replica trick for the simplest
distribution without any correlations as well as for the crosslink
distribution of Deam and Edwards \cite{DeEd76} which is believed to be
more realistic. In {\it both} cases we find a logarithmic divergence
of the viscosity for a network of Brownian particles, as the gelation
transition is approached.

\section{Scaling description}
The above crosslink distribution ignores all spatial correlations and hence
cannot account correctly for the critical behaviour of a three-dimensional
crosslinked polymer melt. To improve on this shortcome
we now assume that the mechanism of crosslinking generates clusters amenable
to the scaling description of percolation. More specifically, we assume 
$\tau_n=n^{-\tau}f\bigl((c_{{\rm crit}}-c)\,n^{\sigma}\bigr)$ with the
scaling function $f$ decaying faster than any polynomial for large arguments
and being virtually constant for small arguments. The critical exponents
$\sigma$ and $\tau$ are related \cite{StCoAd82,StAh94} to the exponents
$\beta$ and $\nu$ by $\sigma^{-1}=d\nu-\beta$ and $\tau=1+d\nu\sigma$. We
recall that the exponents $\beta$, $\nu$ and $\phi$ have been defined in the
introduction, the latter ruling the behaviour of the resistance. 

It has turned out in the preceeding section that crosslinked Brownian particles
and crosslinked polymer chains have identical critical behaviour. We adopt this
simplification by assuming that it holds true in the present case, too.
Using (\ref{Eq9}), we thus get
\begin{equation} \label{Eq13}
  \left\langle \eta \right\rangle_n = 
  \frac{\zeta a^{2}}{4dn^2\tau_{n}} 
  \left\langle 
    \frac{1}{N}\sum_{k=1}^K \delta_{N_k,n}
    \sum_{i,j\in{\cal N}_k}{\cal R}(i;j)
  \right\rangle.
\end{equation}
The second assumption is that $\left\langle\eta\right\rangle_n$, which only
tests cluster of given finite size $n$, does not acquire any irregularity as
the critical point is approached and that
$\left.\left\langle\eta\right\rangle_n\right|_{c=c_{\rm crit}}\sim n^b$ with
some yet unknown exponent $b$. Note that, again, this holds true with $b=1/2$
in the case of completely random crosslinks, see (\ref{Eq11}). Furthermore, we
remark that this assumption can be circumvented, at least to second order in
$\varepsilon=6-d$, by using the more sophisticated methods from
\cite{StJaOe99}. To determine $b$ we use
\begin{equation} \label{Eq14}
  \frac{4d}{\zeta a^{2}}\;\sum_{n=2}^{\infty} n^2
  \tau_n\left\langle\eta\right\rangle_n = 
  \left\langle \sum_{j\in{\cal N}(i)} {\cal R}(i;j) \right\rangle 
  \stackrel{c \uparrow c_{{\rm crit}}}{\sim} 
  (c_{{\rm crit}} - c)^{d\nu - (2/\sigma) -\phi}.
\end{equation}
Here ${\cal N}(i)$ denotes the cluster which contains particle $i$. In the
first step we have used enumeration invariance. In the second step
\cite[eq.~(2.45)]{HaLu87} has been employed. It follows that
$b=\sigma\phi = (2/d_{s}) - 1$ and 
(\ref{Eq10}) finally leads to the scaling relation (\ref{Eq1}) for the averaged
viscosity.
\section{Discussion}
The mean-field approximation \cite{St77} for the percolation
transition reproduces the critical behaviour \cite{ErRe60} of
completely random crosslinks. The value $k=0$ for $d\ge 6$, which is
found from (\ref{Eq1}) with the help of \cite{StAh94}, suggests a
logarithmic divergence of the viscosity in the latter case. This is
exactly what is found from the rigorous calculation.  Moreover, the
results of \cite{LuWa85,StJaOe99} plugged into (\ref{Eq1}) yield the
$\varepsilon$-expansion $k=\varepsilon/6+11\varepsilon^2/1764+{\cal
  O}(\varepsilon^3)$, $\varepsilon:=6-d$. For $d=3$ we use the results
of high precision simulations \cite{GiLo90} in (\ref{Eq1}) and obtain
$\left.\vphantom{\big(}k\right|_{d=3}\approx 0.71$.  In order to
relate (\ref{Eq1}) to the previously suggested scaling relations
mentioned in the Introduction, we note that if $(2/d_{s})-1 =
(2/d_{f})$ were true 
(which in general is not), one would recover de~Gennes' original
proposal $k=2\nu-\beta$. On the other hand, if the Alexander-Orbach
conjecture, $d_{s}=4/3$, were inserted into $\phi$, the resulting
scaling relation 
would be $k=(d\nu - 3\beta)/2$. 
De~Gennes' second porposal $k=s$ together with Kert\'esz' result
$s=\nu -\beta/2$ yields yet a different scaling relation.
This is even more surprising, because Kert\'esz' result is compatible
with the Alexander-Orbach conjecture in $d=2$.
For comparison we list the 
values $\left.\vphantom{\big(}s\right|_{d=3}\approx 0.73$,
$\left.\vphantom{\big(}(2\nu-\beta)\right|_{d=3}\approx 1.35$ of the
competing proposals for Rouse dynamics. The values are taken from
\cite[table~2]{StAh94}. The similarity between the numerical values of
$k$ and $s$ for $d=3$ has to be interpreted as accidental, because for
$d=2$ one knows from duality \cite{Ha90} that $\phi=s$, which yields
with the high precision data from \cite{Gr99} the clearly distinct
numerical values $\left.\vphantom{\big(}k\right|_{d=2}\approx 1.17$
and $\left.\vphantom{\big(}s\right|_{d=2}\approx 1.31$. Thus the
analogy of the viscosity to the conductivity of a mixture of normal
and superconductors is incompatible with the interpretation advocated
here.

The above result $\left.\vphantom{\big(}k\right|_{d=3}\approx 0.71$
agrees well with the experiments of 
 \cite{AdDeDuHiMu81-AdDeDu85,DuDeAdLu87}. On the other
hand it is not compatible with the possibly oversimplifying albeit
attractive proposal \cite{DaLa90,ArSa90} to interpret the wide
variation of the exponent $k$ of the viscosity as the signature of a
splitting of the static universality class of gelation into different
dynamic ones. Since Rouse and Zimm dynamics are considered
\cite{Oo85,DoEd85} to be at the extreme ends of the strength of the
hydrodynamic interaction, the actual value of $k$ should lie in the
range bounded by the results for the respective dynamics. The
divergence of the viscosity in the Zimm model has to be expected
weaker than in the Rouse model, because the longest relaxation time in
the Zimm model grows with a smaller power of the size of the cluster
than in the Rouse model \cite{DoEd85,DaLa90,Sa94}. Since our result
for Rouse dynamics falls well into the range of small exponent values,
we expect that the extension of our analysis to Zimm dynamics will
yield a change for the worse, as far as the agreement with experiment
is concerned. 
One may think of several other effects which may affect the predictions for
$k$ and have not been taken into account by our simple model. 
Excluded-volume interactions are known to be relevant
even for thermostatic questions \cite{Oo85,DoEd85}.
We refer to \cite{Ca85,DaLa90} and references therein where attempts
are made to account for these effects on a heuristic level.
Additionally, one expects entanglement effects to play a vital r\^ole
in stress relaxation \cite{DoEd85}. However, the static viscosity
measures stress relaxation only on the longest time scales in the sol
phase. It has been argued \cite{CoGiRu93,RuZuMcBa90-MiMcYoHaJo98} that
in the regime close to the transition entanglement effects are less
important because of two reasons: first, there are almost no permanent
entanglements in the sense of interlocking loops. Second, the time
scale of a temporary entanglement of two clusters is determined by the
smaller cluster, whereas the dynamics on the longest time scales is
determined by the larger cluster. 

We hope that the analysis of the simple model presented in this Letter
will serve as a first step towards a more sound understanding of
dynamical effects in gelling polymeric sols. Clearly, further
investigations are necessary to embed the variety of well-developed
scaling pictures within an appealing theoretical framework.

\stars 
This work was supported by the DFG through SFB 345. K.B.\ acknowledges
financial support by the DFG under grant No.\ Br 1894/1-1.

\end{document}